\documentclass[aps,prd,reprint,groupedaddress]{revtex4-1}

\usepackage {graphicx}
\usepackage{amssymb}
\usepackage{amsmath}
\usepackage{units}

\begin{document}


\title{Light-Cone QCD Plasma}


\author{H.J. Pirner}
\affiliation{Institute for Theoretical Physics, University of Heidelberg, Germany}

\author{K. Reygers}
\affiliation{Physikalisches Institut, University of Heidelberg, Germany}


\date{\today}

\begin{abstract}
We deduce the maximum-entropy state of partons created in proton-proton and nucleus-nucleus collisions. This state is characterized by a distribution function $n(x,p_\bot)$ depending on the light-cone fraction $x$ and the transverse momentum $p_{\bot}$ of forward or backward particles, respectively. The mean transverse momentum $\langle p_\bot \rangle$ determines the single parameter of the maximum entropy distribution which is constrained by the sum of all light-cone momentum fractions being unity. The total multiplicity is related to the transverse area of the colliding Lorentz-contracted hadrons. Assuming parton-hadron duality we can compare the model to data from RHIC and LHC.
\end{abstract}


\maketitle

\section{Introduction}
A great challenge in strong-interaction physics has been to relate high-energy scattering experiments to the state of the early universe. Especially, ultra-relativistic heavy-ion collisions promise to create a state of hot and dense matter similar to the quark-gluon plasma \cite{ Rischke:2003mt,Harris:1996zx}  which dominated the early universe approximately one microsecond after the Big Bang. The underlying hypothesis is that in nucleus-nucleus collisions at central rapidities a hot fireball is created with a temperature which is higher than the phase transition or cross-over temperature $T \approx \unit[160]{MeV}$ from a hadronic gas to the quark-gluon plasma. Note, in this  fireball the longitudinal and transverse momenta are equilibrated. Owing to the extended size of the nuclei and the large change in entropy between the quark-gluon plasma and the hadronic gas, the hot matter created in the collision has been considered as a macroscopic state which lives long enough to determine the  main features of the collision. This scenario can in principle also be related to the rapidity distribution of the produced particles if a series of fireballs spread out along the rapidity axis \cite{Schnedermann:1993ws,BraunMunzinger:1994xr,Becattini:2007ci} is constructed. 

In this letter we propose a different picture based mostly on experimental information. Our goal is to describe the inclusive cross section of charged particles produced in proton-proton (pp) collisions 
\begin{align}
\frac{dN}{dy d^2 p_{\bot}}&=\frac{dN}{d\ln x d^2 p_{\bot}}\\
                          &=\frac {d \sigma}{ \sigma_{in} dy d^2 p_{\bot}}
\end{align}
on the basis of a statistical distribution function which maximizes the entropy of the produced partons given certain constraints. We consider the light-cone momenta of the partons with energies $\epsilon$ and longitudinal momenta $p_{z}$ relative to the light-cone momentum of the incoming proton with $(E,P_{z},0)$:
\begin{equation}
x= \frac{\epsilon+p_{z}}{E+P_{z}} = \frac{p_+}{P_+}.
\end{equation}

On the basis of light-cone momentum conservation, we then determine the maximum-entropy distribution \cite{Jaynes:1957zza,Jaynes:1957zz} for a given transverse energy which we call light-cone plasma distribution. Our approach aims at a macroscopic description of the very soft part of the multiplicity distribution in hadronic collisions using as little dynamical input as possible.  We refer to the literature for microscopic calculations of rapidity distributions based on parton shower Monte Carlo event generators \cite{Webber:2011} or based on unintegrated gluon distributions from saturation models \cite{Kharzeev:2004if}.
   
The light-cone plasma distribution is a new state of matter, different from the thermal quark-gluon plasma distribution in the early universe. It agrees rather well with data when we assume a gradual transition from partons to hadrons, i.e., when we use parton/hadron duality to relate our model to the measured rapidity distributions.   

In the following we consider symmetric collision partners and discuss the produced partons separately in the forward and backward hemisphere. We limit ourselves to gluons only, i.e., to Bose statistics. It is important to emphasize the light-cone property of the maximum-entropy distribution. The dynamics of collisions at high energies is governed by a light-cone Hamiltonian which is boost invariant and determines wave functions depending on transverse momentum and light-cone fractions which label the eigenstates. The density matrix resulting after the collision can be built up from an incoherent mixture of such multi-parton states. Without a boost invariant formulation one cannot define a number density of partons since in each reference system it will be different. In the rest system of the proton the gluons will all sit in the strings holding the quarks together, whereas in fast-moving proton the virtual gluons materialize as partons carrying a sizable momentum fraction.  Light-cone physics has been very successful in determining hard exclusive processes in high-energy collisions \cite{Lepage:1980fj}. We think that it is also important to describe soft inclusive cross-sections via the maximum-entropy principle.

The outline of the paper is as follows. In section 2 we give a theoretical motivation for the light-cone plasma distribution. In section 3 we calculate the parameters of the light cone distributions for pp collisions at $\sqrt{s}=\unit[200]{GeV}$ and \unit[7000]{GeV}. In section 4 we consider nucleus-nucleus collisions as a superposition of pp collisions where the mean transverse momentum of the partons is broadened. Section 5 is devoted to a comparison of the entropy of the light-cone distribution and the conventional smeared fireball distribution. Section 6 gives our conclusions.

\section{Motivation for the Light Cone Distribution}
The  multiplicity distribution for partons can be motivated  from the maximum-entropy principle. For this purpose we have to find the phase space for a system of partons which move on the light cone. The entropy of the partonic system is then proportional to the logarithm of the integrated phase space. This entropy has finally to be maximized given certain constraints which take into account the conservation of light-cone momentum and the limited transverse energy produced in hadronic collisions. 

On the light cone, phase space includes the transverse spatial coordinate $b_{\bot}$, the transverse momentum $p_{\bot}$, the longitudinal light-cone momentum $p_+$ and the longitudinal spatial variable $x_-=1/2(x_0-x_3)$. These variables are handled like in conventional thermodynamics, i.e., the phase space multiplied by the gluon degeneracy factor $g=2 (N_c^2-1)$ and divided by the Planck constant $h^3=(2 \pi \hbar)^3$  gives the number of available quantum states $G$:
\begin{align}
G_{b_{\bot},p_{\bot},p_+,x_-}&=g  \frac{d^2 b_{\bot} d^2p_{\bot}dp_+dx_-}{(2 \pi)^3} \\
                                  &=g  \frac{d^2 b_{\bot} d^2p_{\bot}}{(2 \pi)^2} dx \frac{d \rho}{2 \pi}.
\end{align}
For high energies, Feynman scaling is  a good phenomenological concept, therefore  we have multiplied and divided this expression by $P_+=E+P_z$ to obtain the light-cone momentum $x=p_+/P_+$ and the longitudinal light-cone distance $\rho=x_-P_+$ which are canonically conjugate variables. We refer to ref. \cite {Pirner:2009zz} to show the role of $\rho$ in the light cone Hamiltonian of a meson built from a valence quark and antiquark. We further make the simplifying assumption that the distribution function and consequently the entropy are homogeneously distributed in transverse space. The integration over the $b_{\bot}$-coordinate can then be executed and gives the area $L_{\bot}^2$. 

An estimate of the integral of the scaled light-cone distance $\int \frac{d \rho}{2 \pi}$ is more subtle, since it is not independent on the rest of the variables. It has to be done separately for valence and sea partons. Let us describe the fast-moving system by a Lorentz-factor $\gamma \rightarrow \infty$. Then the valence quarks occupy a decreasing longitudinal extension $\Delta x_- \approx \frac {L_z}{\gamma}$ whereas the extension of the sea partons remains fixed.  Consequently, the longitudinal distance for valence quarks scaled with the increasing light-cone momentum $P_+$ of the proton is constant and yields for $m=\unit[0.938]{GeV}$ a factor of $O(1)$:
\begin{equation}
\int \frac{d \rho_\mathrm{val}}{2 \pi} \approx \frac {L_z}{\gamma} \frac{m \gamma}{2 \pi}\approx 1 .
\end{equation}

For sea partons with $x \rightarrow 0$ the scaled distance diverges, cf. \cite{Roberts:1990ww}:
\begin{equation}
\int \frac{d \rho_\mathrm{sea}}{2 \pi} \approx  \frac{ 1}{x P_+}P_+\rightarrow \infty.
\end{equation}

We interpolate the $x$ dependence of these two limiting cases for the integral over the scaled distance in the following way:
\begin{equation}
\int \frac{d \rho}{2 \pi} \approx \frac{1}{x}.
\end{equation}
A possible pre-factor of $1/x$ cannot be determined more accurately and has to be absorbed into the transverse area  $L_{\bot}^2$. The so motivated ansatz for the phase space on the light cone is crucial for all further derivations. It deviates from the flat measure by the factor $1/x$:
\begin{equation}
G_{x,p_{\bot}}=g L_{\bot}^2 \frac{ d^2 p_{\bot}}{(2 \pi)^2} \frac{dx}{x}.
\end{equation}

Gluons are bosons; therefore, they can occupy the phase-space cells in multiples. The binomial of the combined number of particles and states over the number of states gives the number of possibilities to distribute $N_{x,p_{\bot}}$ gluons, i.e., bosons, on $G_{x,p_{\bot}}$ quantum states: 
\begin{equation}
\label{eq:delta_gamma}
\Delta \Gamma_{x,p_{\bot}}= \frac{(G_{x,p_{\bot}}+N_{x,p_{\bot}}-1)!}{(G_{x,p_{\bot}}-1)! N_{x,p_{\bot}}!}.
\end{equation}

The entropy of the system is defined by the logarithm of the phase space. In Eq. \ref{eq:delta_gamma} we use Stirling's formula and set $G_{x,p_{\bot}}-1 \approx  G_{x,p_{\bot}}$ for large numbers of quantum states and particle numbers. Then one gets the entropy from the summation of the individual phase space elements:
\begin{align}
\label{eq:entropy1}
S &=\sum \ln(\Delta \Gamma_{x,p_{\bot}})\\
\label{eq:entropy2}
  &=\sum G_{x,p_{\bot}}[(1+n_{x,p_{\bot}}) \ln(1+n_{x,p_{\bot}})-n_{x,p_{\bot}} \ln n_{x,p_{\bot}}]  
\end{align}
where the mean occupation number of each quantum state is defined as
\begin{equation}
n_{x,p_{\bot}} = \frac{N_{x,p_{\bot}}}{G_{x,p_{\bot}}}.
\end{equation}

In high-energy collisions the searched-for maximum entropy distribution has to satisfy the following two requirements:
\begin{align}
\label{eq:req1} 
\sum G_{x,p_{\bot}}x\,n_{x,p_{\bot}}&=1\\
\label{eq:req2}
\sum G_{x,p_{\bot}} \,p_{\bot} n_{x,p_{\bot}}&=\langle E_{\bot} \rangle.
\end{align}

The first constraint means that the $x$ fractions of all partons emitted in the positive hemisphere add up to unity, i.e., their light-cone momenta equal the light-cone momentum of the parent proton. The second constraint defines the total transverse energy released in the collision in the positive hemisphere. These constraints are added with Lagrange parameters $1/\lambda$ and $w$ to the entropy above. These two constraints are sufficient to  determine the $x, p_{\bot}$ dependence of the distribution function since we have fixed the distribution in coordinate space. The resulting functional $S+1/\lambda \cdot \langle E_{\bot} \rangle+w \cdot 1$ can then be varied with respect to $n_{x,p_{\bot}}$ to obtain the maximum entropy density:
\begin{equation}
\frac{\delta (S+ \frac{1}{\lambda}\sum p_{\bot} n_{x,p_{\bot}}+w \sum x n_{x,p_{\bot}})}
{\delta n_{x,p_{\bot}}}=0.
\end{equation}
 
By choosing the cell sizes small we convert the sums into integrals over the continuum variables $x,p_{\bot}$ and the discrete distribution $n_{x,p_{\bot}}$ becomes $n(x,p_{\bot})$, the light-cone plasma distribution function
\begin{equation}
n(x,p_{\bot}) = \frac{1}{e^{\frac{p_{\bot}}{\lambda}+x w }-1}.
\end{equation}

The light-cone plasma distribution together with the measure generates a distribution for the Lorentz-invariant yield of the form
\begin{equation}
\frac{d N}{dy d^2p_{\bot}}= \frac{g L_{\bot}^2}{(2 \pi)^2} 
\frac{1}{\exp\left[p_{\bot}\left(\frac{1}{\lambda}+\frac{w e^{|y|}}{\sqrt{s}}\right)\right]-1}.
\end{equation}

This is the connection of the maximum-entropy distribution on the light cone with the semi-inclusive cross section. The factor $1/(2\pi)^2$ has its origin in the transverse phase space cell $\frac{d^2b_{\bot} d^2 p_{\bot}}{\hbar^2 (2 \pi)^2}$.  The distribution is consistent with the two constraints of light-cone momentum conservation and total transverse energy which have the following form in continuum variables. The $x$ integration is executed in one hemisphere in the cm system, where the partons released by the proton projectile or target respectively are to be found:
\begin{eqnarray}
\label{eq:lcm}
g  L_{\bot}^2\int \frac{d^2 p_{\bot}}{(2 \pi)^2} \int \frac{dx}{x} 
x\,n(x, p_{\bot})&=&1\\
\label{eq:et}
g  L_{\bot}^2\int \frac{d^2 p_{\bot}}{(2 \pi)^2} \int \frac{dx}{x} 
p_{\bot} \, n(x, p_{\bot})&=&\langle E_{\bot} \rangle.
\end{eqnarray}

The phenomenological description of the multiplicity distribution has three parameters $L_{\bot}^2$, $\lambda$, and $w$. The parameter $\lambda$ plays the role of an effective transverse ``temperature''. The ``softness'' $w$ is related to the mean $x$. With increasing center-of-mass energies we expect that the effective transverse temperature $\lambda$ and the softness $w$ increase: The collision becomes ``hotter'' and the particle distributions ``softer''.  The effective transverse temperature $\lambda$ is calculated from the mean transverse momentum which is equal to the ratio of the transverse energy and multiplicity in one hemisphere:
\begin{equation}
\langle p_\bot \rangle = \langle E_{\bot} \rangle/( N/2 )
\end{equation}
with
\begin{equation}
N/2=g L_{\bot}^2\int \frac{d^2 p_{\bot}}{(2 \pi)^2} \int \frac{dx}{x} n(x,p_{\bot}).
\end{equation}
For a given $L_{\bot}$, $\lambda$, and $w$ the multiplicity is uniquely defined when the cut on the $x$ integration is taken as $x_\mathrm{min}=\sqrt{p_{\bot}^2+m_{\pi}^2}/\sqrt{s}$. We describe in the next section how this distribution fits the data.

\section{Parameters of Light-Cone Distributions in pp Collisions}
When we want to relate the theoretical distribution of the gluon plasma to multiplicities of produced particles we must apply a simplified form of parton-hadron duality. We assume that all particles are pions and replace transverse momentum by the transverse mass. The interpretation of the pre-factor $g L_{\bot}^2$ for pions has to be changed. For pions a smaller degeneracy factor $g_{\pi}=3$ corresponds to a larger area $L_{\bot,\pi}^2=\frac{16}{3} L_{\bot}^2$ at freeze out. The pion  multiplicity distribution then has the following form:
\begin{equation}
\frac{d N}{dy d^2p_{\bot}}= \frac{g L_{\bot}^2}{(2 \pi)^2} 
\frac{1}{\exp\left[m_{\bot}\left(\frac{1}{\lambda}+\frac{w e^{|y|}}{\sqrt{s}}\right)\right]-1}
\end{equation}
with 
\begin{equation}
m_{\bot}=\sqrt{p_{\bot}^2+m_{\pi}^2}.
\end{equation}

For not too large rapidities, its integral over transverse momentum can be expanded in powers of $m_{\pi} a$: 
\begin{equation}
\frac{d N}{dy} \approx \frac{ \pi g L_{\bot}^2 \lambda^2}{12  
\left( 1+ w \lambda \frac{e^{|y|}}{\sqrt{s}}\right)^2}  \left(1-\frac{6 m_{\pi} a}{\pi^2}+\frac{3 m_{\pi}^2 a^2}{2 \pi^2} - ...\right)
\label{eq:dndy}
\end{equation}
with
\begin{equation}
a=\frac{1}{\lambda}+ w \frac{e^{|y|}}{\sqrt{s}}.
\end{equation}
Collider experiments preferentially take data around central rapidity as a function of the pseudorapidity $\eta$ which requires only to measure the angle of each particle relative to the beam axis. Since we saturate the reaction products by pions, charged hadrons make up $2/3$ of the total multiplicity:  
\begin{equation}
\frac{d N_\mathrm{ch}}{d\eta d^2p_{\bot} } =\frac{2}{3} 
\sqrt{1- \frac{m_\pi^2}{m_\bot^2 \cosh^2 y}} 
\frac{d N}{dy d^2p_{\bot}}.
\end{equation}

In \cite{Khachatryan:2010us} the $\sqrt{s}$ dependence of $dN_\mathrm{ch}/d\eta|_{\eta=0}$ and $\langle p_T \rangle_{\eta=0}$ were parameterized as
\begin{equation}
dN_\mathrm{ch}/d\eta|_{\eta=0} = 2.716 - 0.307 \ln s + 0.0267 \ln^2s
\end{equation}
and 
\begin{equation}
\langle p_T \rangle_{\eta=0} = 0.413 - 0.0171 \ln s + 0.00143 \ln^2s.
\end{equation}
We use these parameterizations as experimental inputs. The theoretical constraint of light-cone momentum conservation (cf. Eq.~\ref{eq:req1} or \ref{eq:lcm}) serves as third input to determine the three unknown parameters $L_{\bot}, \lambda$, and $w$ of the light-cone plasma distribution. Unlike in $e^+e^-$-collisions it cannot be expected that the total center-of-mass energy is available for particle production. This may be described by subtracting from the cm energy the energy of the leading particles $E_\mathrm{leading}$, cf.~\cite{Back:2006yw}. Therefore, we will represent results with a $K$ factor which reduces the effective cm energy  and is defined as
\begin{equation}
K \sqrt{s}=\sqrt{s}-2 \langle E_\mathrm{leading} \rangle.
\end{equation} 
A realistic consideration of the gluons which do not alone carry the light-cone momentum of the proton would go in the same direction.

\begin{figure}
\centering
\includegraphics[width=\linewidth]{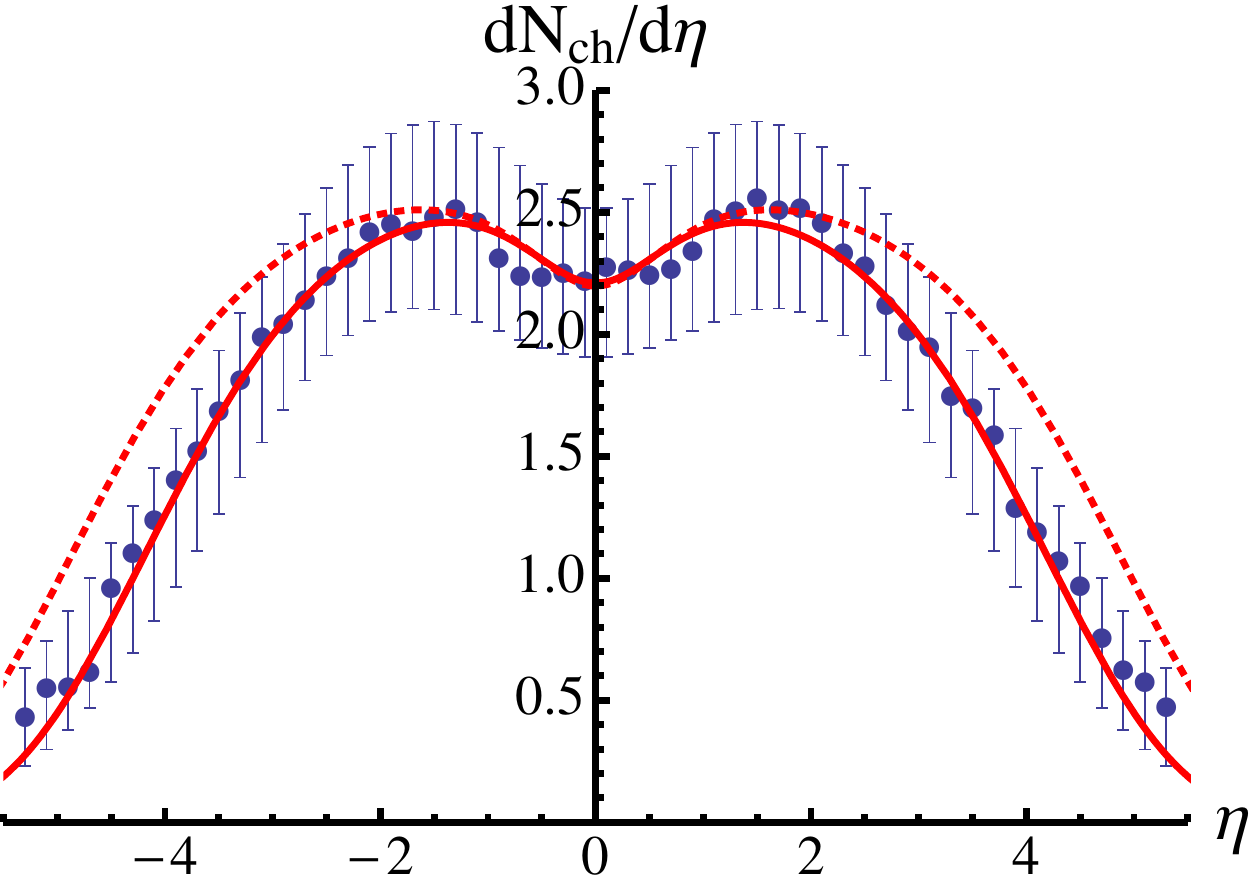}
\caption{Data points show the charged-particle pseudorapidity distribution in p+p collisions at $\sqrt{s}= \unit[200]{GeV}$ 
from \cite{Alver:2010ck}. The curves represent the light-cone plasma distributions (solid line $K=0.5$, dotted line $K=1.0$).}
\label{fig:dnchdeta_pp_200gev}
\end{figure}
\begin{figure}
\centering
\includegraphics[width=\linewidth]{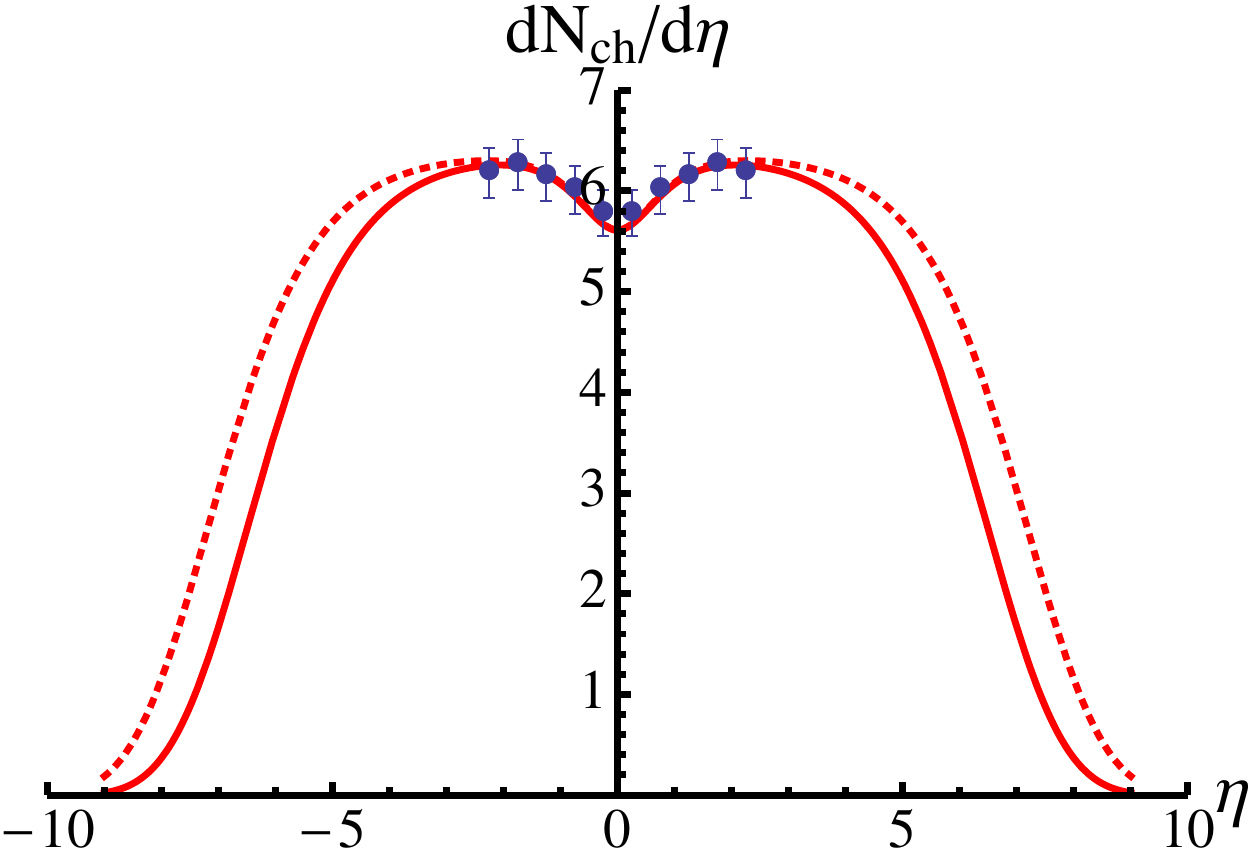}
\caption{Data points show the charged-particle pseudorapidity distribution in p+p collisions at $\sqrt{s}= \unit[7000]{GeV}$ from \cite{Khachatryan:2010us}, the curves represent the light-cone plasma distributions (solid line: $K=0.5$, dotted line $K=1.0$).}
\label{fig:dnchdeta_pp_7000gev}
\end{figure}
Figures~\ref{fig:dnchdeta_pp_200gev} and \ref{fig:dnchdeta_pp_7000gev}  show the comparison of the light-cone plasma distribution with the data from RHIC \cite{Alver:2010ck} and LHC \cite{Khachatryan:2010us}. One sees that the light-cone plasma distribution for $K=0.5$ can reproduce the main features of the data quite well. In Table~\ref{tab:pars} we have compiled the parameters $L_{\bot},\lambda$, and $w$ determined from $dN_\mathrm{ch}/d\eta|_{\eta=0}$ and the mean transverse momentum $\langle p_T\rangle$ at $\eta=0$ for these energies. From the table one can see that the increase of the rapidity distribution at $\eta=0$ is of order $\lambda^2$ (cf. Eq.~\ref{eq:dndy}).
\begin{table}[h]
\begin{tabular}{c|c|c|c|c|c|c}
$\sqrt{s}$ & $L_\bot$ & $\lambda$ & $w$ & $dN_\mathrm{ch}/d\eta|_{\eta=0}$ & $\langle p_T|_{\eta=0} \rangle$ \\
(TeV)         & (fm) & (GeV)         &         &         & (GeV)                          &                 \\
\hline
0.20 & 1.34 & 0.183 & 3.44 & 2.20 & 0.39  \\
0.90 & 1.25 & 0.216 & 5.36 & 3.48 & 0.45  \\
2.76 & 1.28 & 0.252 & 6.81 & 4.56 & 0.50  \\
7.00 & 1.20 & 0.288 & 8.21 & 5.65 & 0.56 
\end{tabular}
\caption{For different cm energies and $K=0.5$ we present the size parameter $L_{\bot}$, the effective transverse temperature $\lambda$, and the softness $w$ of the light-cone distributions. The following columns show the experimental input: the multiplicity $dN_\mathrm{ch}/d\eta$ and the mean transverse momentum $\langle p_T\rangle$ at pseudorapidity $\eta=0$.}
\label{tab:pars}
\end{table}

\begin{figure}
\centering
\includegraphics[width=\linewidth]{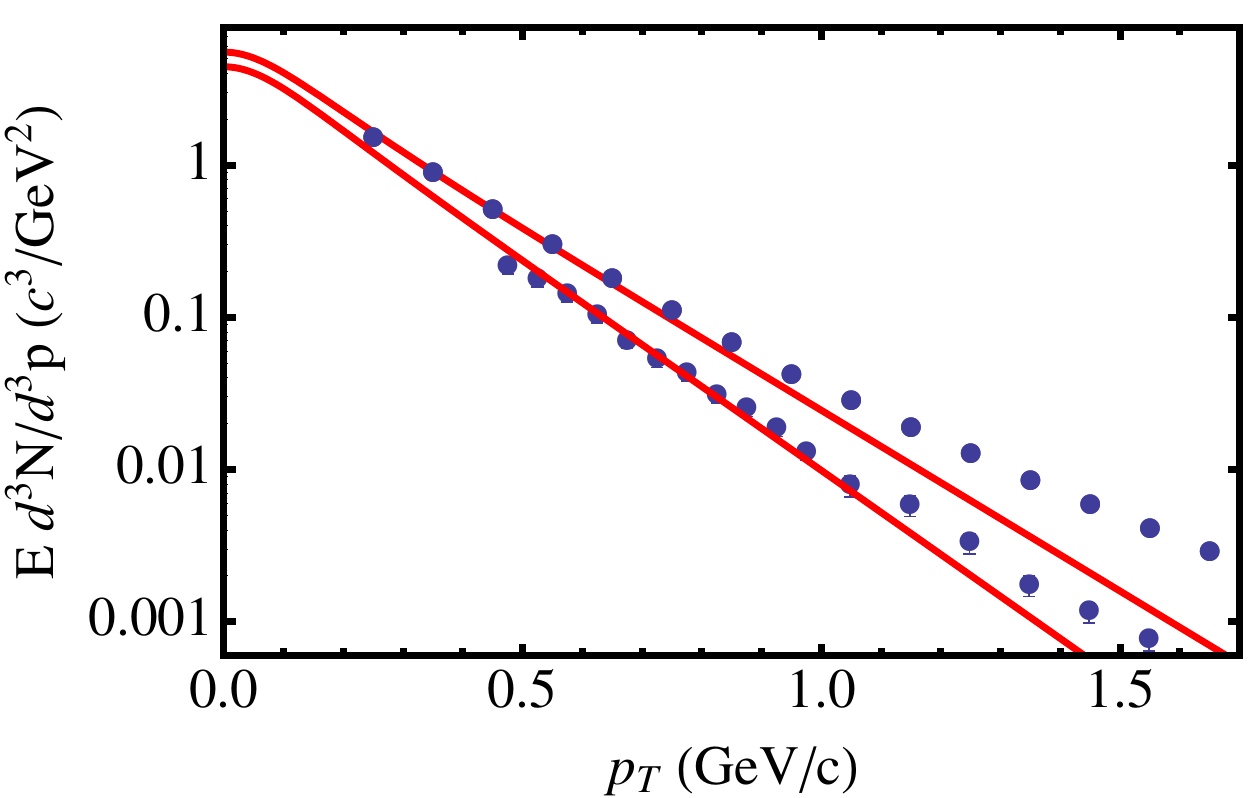}
\caption{Transverse momentum dependence of the inclusive cross section at $\sqrt{s}=\unit[200]{GeV}$ for charged hadrons ($(h^++h^-)/2$) at $\eta=0$ (upper points) \cite{Albajar:1989an} and positive pions at $\eta=3.3$ (lower points) \cite{Arsene:2007jd}. The respective plasma distributions for $K=0.5$ are shown as full lines.}
\label{fig:dnchdpt_pp_200gev}
\end{figure}

A further test of the light-cone plasma distribution is given by a measurement of the multiplicity distribution as a function of transverse momentum for different rapidities. In Fig.~\ref{fig:dnchdpt_pp_200gev} we plot two experimental transverse momentum spectra for charged hadrons ($(h^++h^-)/2$) at $\eta=0$ (upper points) \cite {Albajar:1989an} and for positive pions at $\eta=3.3$ (lower points) \cite{Arsene:2007jd} at the same cm energy of $\sqrt{s} = \unit[200]{GeV}$. The full drawn curves represent the corresponding light-cone plasma distributions at these rapidities. They fit the inclusive cross sections up to \unit[1]{GeV}/$c$ rather well. For higher momenta significant contributions from hard scattering are expected. The plasma distribution describes the fall-off of the cross sections with transverse momentum by an effective transverse temperature $\lambda_\mathrm{eff}(y)$ which depends on rapidity $y$:
\begin{equation}
\frac{d N}{dy d^2p_{\bot}}= \frac{g L_{\bot}^2}{(2 \pi)^2} 
\frac{1}{\exp\left[m_{\bot}/\lambda_\mathrm{eff}(y)\right]-1}
\end{equation}
with
\begin{equation}
\lambda_\mathrm{eff}(y)=\frac{\lambda}{1+w \lambda\exp |y|/\sqrt{s}}.
\end{equation}
Due to light-cone momentum conservation the effective temperature decreases with increasing rapidity. Therefore, the transverse momentum spectra fall off faster at larger rapidities. It would be good to test the light-cone plasma distribution over a larger domain in $y$ or $\eta$ and $p_{\bot}$.

\section{Nucleus-Nucleus Collisions}
It is possible to extend the parametrization of the multiplicity distributions to nucleus-nucleus collisions. We can use the universality of the light-cone plasma distribution originating from pp collisions. In A-A collisions, we multiply the underlying pp multiplicity by the number of participating nucleons $N_\mathrm{part}$ and take into account the increase of $\langle p_{\bot} \rangle$ with centrality: 
\begin{equation}
\frac{d N_\mathrm{ch}^{AA}}{d\eta d^2p_{\bot} } =\frac{N_\mathrm{part}}{2}\frac{2}{3} 
\sqrt{1 - \frac{m_\pi^2}{m_\bot^2 \cosh^2 y}} 
\frac{d N(\langle p_{\bot} \rangle)}{dy d^2p_{\bot}}.
\label{eq:npart_scaling}
\end{equation}

In principle, colliding rows of nucleons contain varying numbers of nucleons in projectile and target nucleus which may lead to a small shift in the total cm rapidity and total cm energy. But these corrections are minor kinematic corrections in nucleus-nucleus (A-A) collisions. In p-A collisions, however, the kinematics of the different row configurations is expected to be more important. In both p-A and A-A collisions the increase of the mean transverse momentum $\langle p_{\bot} \rangle$ with the number of participants is an important feature of the collision which has to be taken into account. The initial parton distributions in the projectile nucleus will be broadened by the interaction with the nucleons in the target nucleus and vice versa. 
\begin{figure}[!]
\centering
\includegraphics[width=\linewidth]{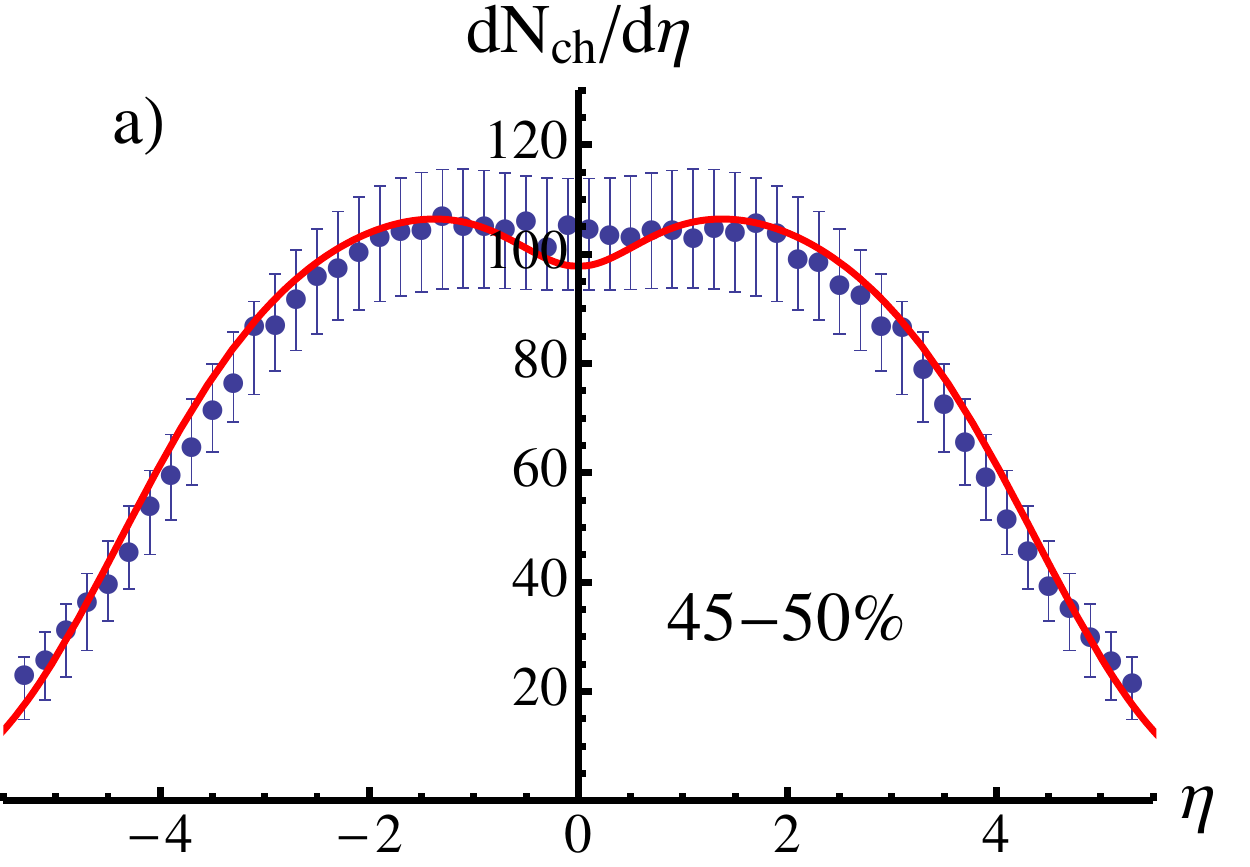}
\includegraphics[width=\linewidth]{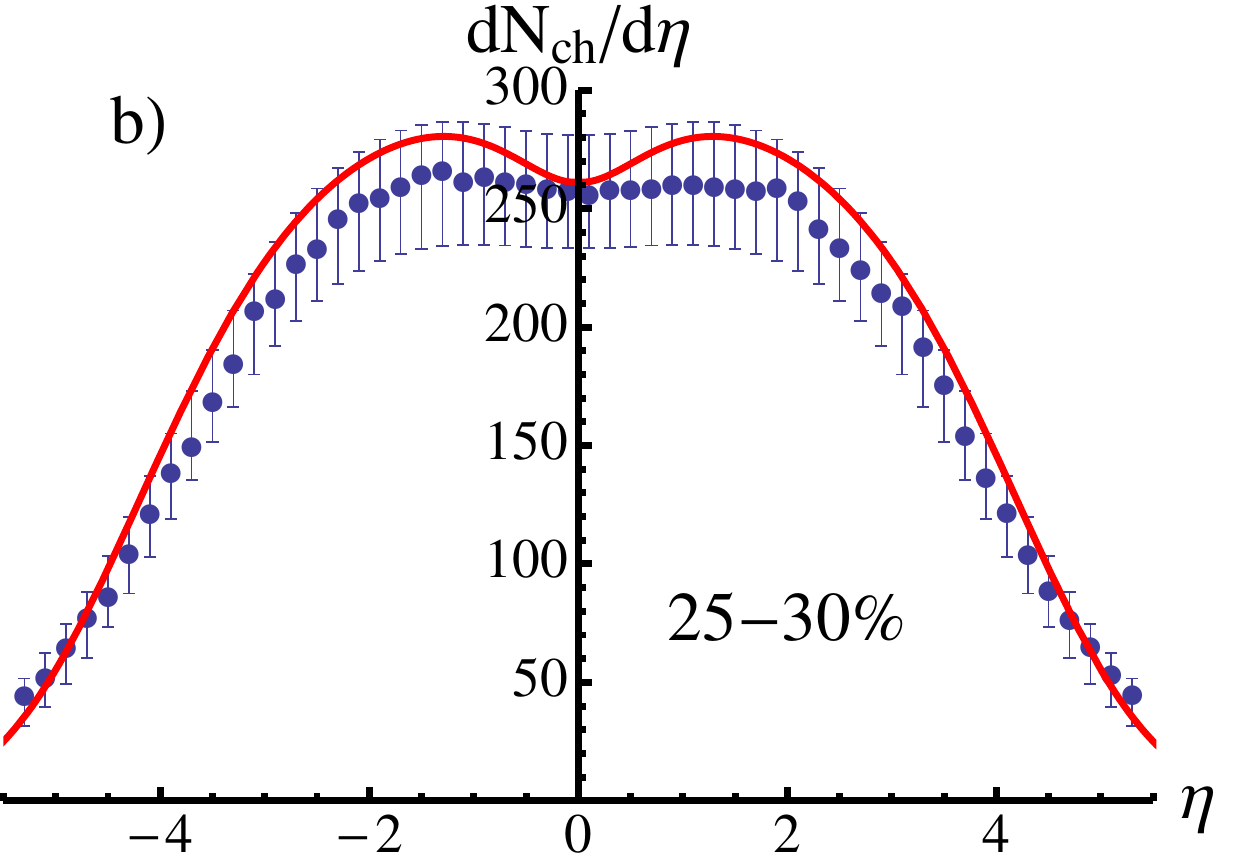}
\includegraphics[width=\linewidth]{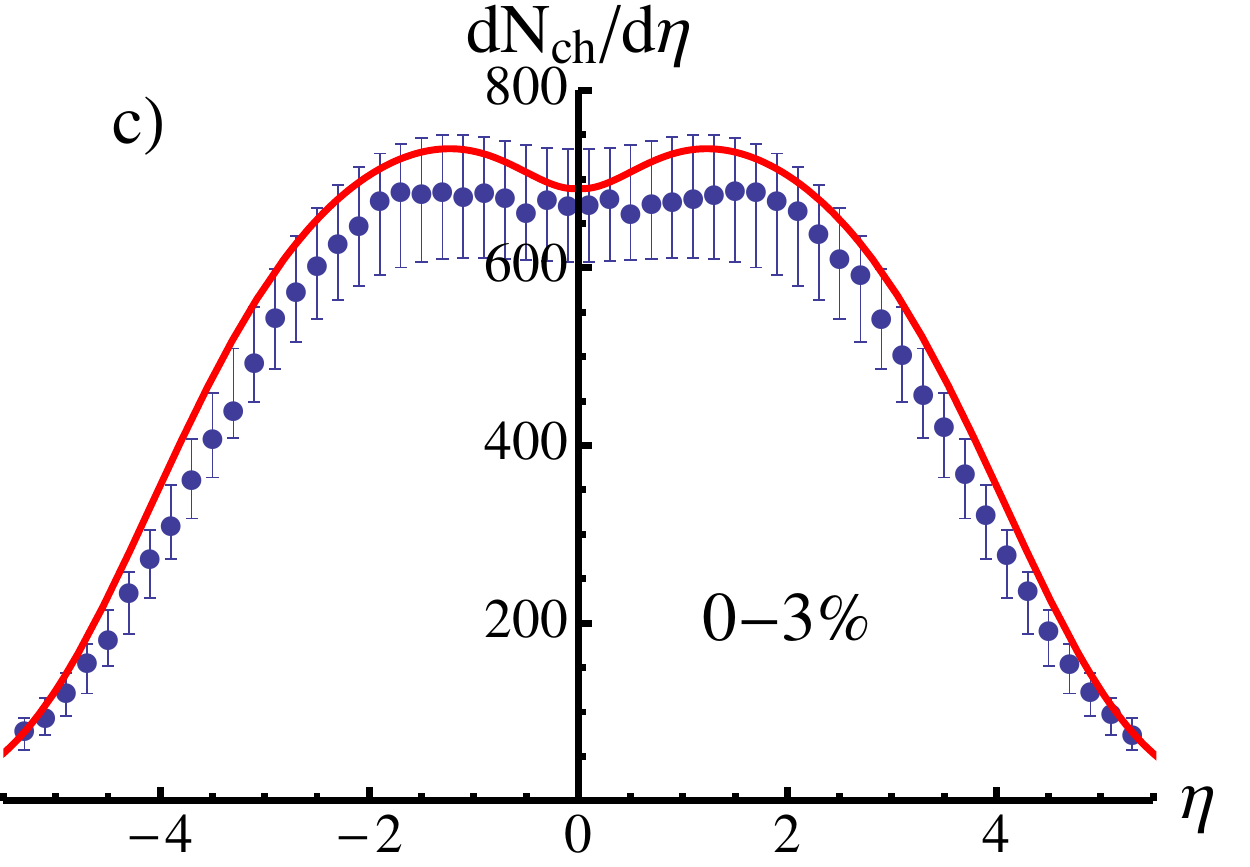}
\caption {Charged particle pseudorapidity distributions from Phobos \cite{Alver:2010ck} in Au+Au collisions at $\sqrt{s_{NN}} = \unit[200]{GeV}$ for the following centrality classes: a) $45-50\,\%$, b) $25-30\,\%$, and c) $0-3\,\%$. Data are compared with the theoretical curves obtained from the light-cone plasma distributions ($K=1$)}
\label{fig:dnchdeta_heavy_ions}
\end{figure}

We parameterize the $\langle p_T \rangle$ data from the STAR experiment \cite{Ullrich:2003az} as
\begin{equation}
\langle p_T \rangle|_{\eta=0} =  p_0 + (p_1 - p_0) (1 - \exp(-(N_\mathrm{part} - 2)/p_2))
\end{equation}
with
\begin{align}
p_0 &= \unit[0.390]{GeV}/c\\
p_1 &= \unit[0.516]{GeV}/c\\
p_2 &= 52.
\end{align}
We will show in a separate publication how the observed broadening can be estimated by considering the multiple scattering of partons in pp collisions using the thickness and density \cite{Domdey:2008aq} of the other nucleus. Here we use the observed experimental transverse momentum broadening \cite{Ullrich:2003az} to fit the input $\lambda$ and $w$ values, cf. Table~\ref{tab:au_au}. Note that for increasing mean transverse momentum the softness $w$ increases because of the light-cone momentum sum rule. In Table~\ref{tab:au_au} we show for each centrality 
the $\langle p_{\bot} \rangle$ of charged hadrons \cite{Ullrich:2003az} and the scaled central rapidity density \cite{Alver:2010ck}. To determine the light-cone plasma parameters $\lambda$ and $w$ we fix $L_{\bot} = \unit[1.12]{fm}$. 
\begin{table}[h]
\begin{tabular}{c|c|c|c|c|c}
centrality       &$N_\mathrm{part}$ & $\langle p_T \rangle$ & $\frac{1}{N_\mathrm{part}/2}\frac{dN_\mathrm{ch}^{AA}}{d\eta}$ & $\lambda$ & $w$  \\
                &                            & (GeV)          &         & (GeV)       &                            \\                 
\hline
45-50\% &  65   & 0.478 &  2.96 & 0.238 & 4.516\\
25-30\% & 150  & 0.509 &  3.63 & 0.253 & 5.184\\
0-3 \%    & 361  & 0.516 &  3.81 & 0.263 & 5.650
\end{tabular}
\caption{For $\sqrt{s}=\unit[200]{GeV}$ Au+Au collisions the table gives the centralities, the number of participants $N_\mathrm{part}$, mean transverse momentum $\langle p_T \rangle$, and $dN_\mathrm{ch}^{AA}/d\eta/(N_\mathrm{part}/2)$ at $\eta=0$ together with the resulting light-cone plasma parameters for $K=1$.}
\label{tab:au_au}
\end{table}

\begin{figure}
\centering
\includegraphics[width=\linewidth]{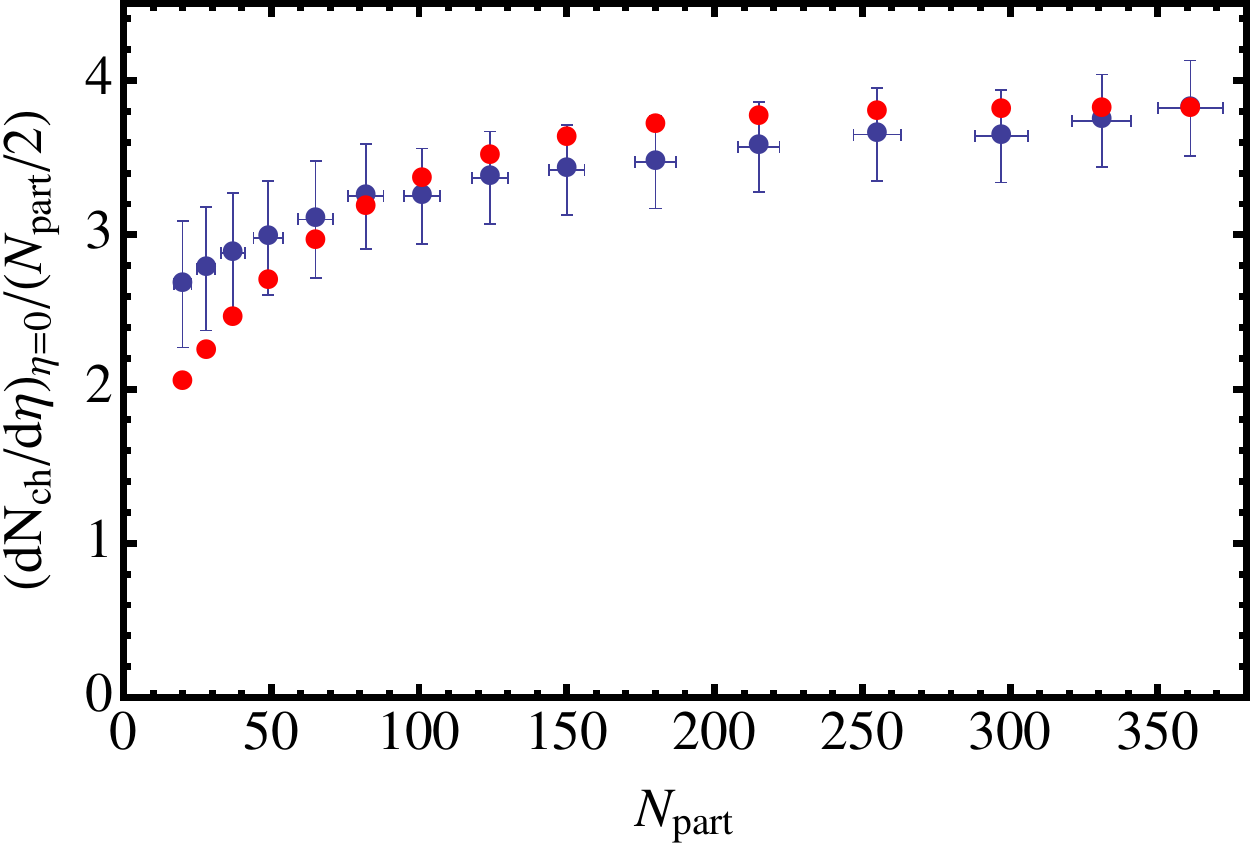}
\caption{The experimental (blue points) \cite{Alver:2010ck} and theoretical (red points, $K=1$ for all centralities) scaled central multiplicities in Au+Au collisions at $\sqrt{s_{NN}} = \unit[200]{GeV}$ are shown as a function of the number of participants.}
\label{fig:dnchdeta_vs_npart}
\end{figure}

The theoretical rapidity distributions in Figs.~\ref{fig:dnchdeta_heavy_ions} a, b, c reproduce the variation of the measured rapidity distributions rather well for $K=1$. Since for fixed energy the size of the nucleon-nucleon overlap area $L_{\bot}^2$ is constant the increase of $dN_\mathrm{ch}^{AA}/d\eta$ at $\eta=0$ is due to the number of participants and $\lambda^2$, cf. Eq.~\ref{eq:dndy}, \ref{eq:npart_scaling}. The central multiplicity divided by the number of participants (cf. Fig.~\ref{fig:dnchdeta_vs_npart}) illustrates the increase originating from the higher mean transverse momentum or effective transverse temperature $\lambda$.

\begin{figure}
\centering
\includegraphics[width=\linewidth]{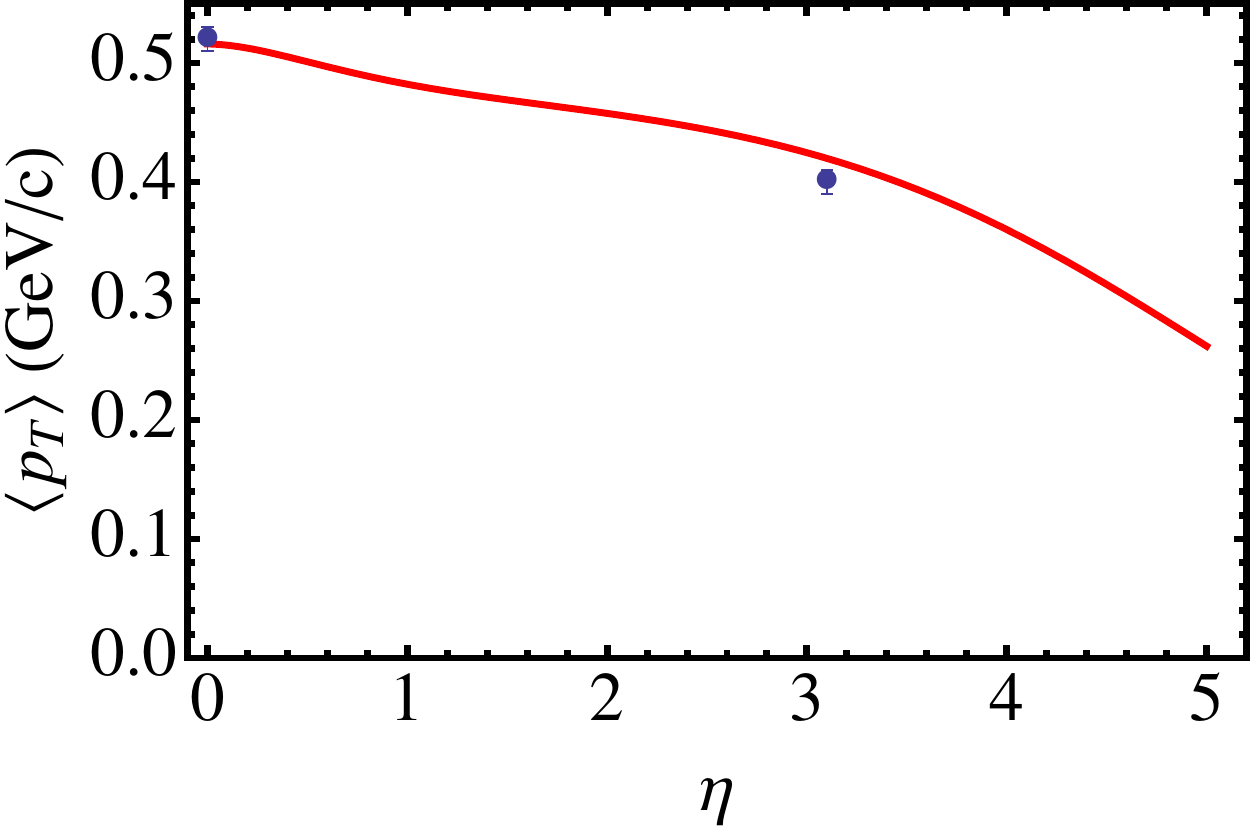}
\caption{Mean $p_T$ as a function of $\eta$ in central Au+Au collisions at $\sqrt{s_{NN}}=200$~GeV compared to data from STAR \cite{Simon:2004xk}.}
\label{fig:meanpt}
\end{figure}

In Fig.~\ref{fig:meanpt} we give the dependence of the mean transverse momentum on rapidity for $\sqrt{s_{NN}} = \unit[200]{GeV}$. The data points are from STAR \cite{Simon:2004xk} and the the curve is  calculated for central Au-Au collisions. As shown before for pp collisions the conservation of light-cone momentum makes the effective transverse temperature decrease for larger rapidities.

\section{Entropy of the Light-Cone Plasma and the Thermal Plasma}
It is instructive to compare the light-cone plasma distribution with a sum of thermal distributions boosted along the $z$ axis 
\cite{Schnedermann:1993ws,BraunMunzinger:1994xr,Becattini:2007ci}. Thereby one mimics the wide rapidity plateau seen in data: 
\begin{eqnarray}
\frac{dN_0^\mathrm{bt}(y,p_{\bot})}{dy d^2p_{\bot}}=\frac{g L_{\bot}^2 L_z}{(2 \pi)^3} \frac{E_p}{e^{E_p/T}-1}\\
\frac{dN^\mathrm{bt}(y,p_{\bot})}{dy d^2p_{\bot}}=\int_{-y_\mathrm{max}}^{y_\mathrm{max}}du \frac{dN_0^{bt}(y-u,p_{\bot})}{dy d^2p_{\bot}}.
\end{eqnarray}
For a comparison of entropies the same transverse area of the system ($L_\bot=\unit[1.12] {fm}$) has to be chosen and the same constraints have to be included for both distributions. As constraints we use the mean transverse energy, the multiplicity, and the light-cone momentum sum rule. We take as an example the light-cone distribution for pions corresponding to $\sqrt{s_{NN}}=\unit[200]{GeV}$ central Au-Au collisions and compare it with a the boosted thermal inclusive particle distribution. Both distributions are divided by the number of participants.

The thermal spectra at fixed rapidity have a rather different functional form, but we can fit the temperature to reproduce the transverse energy. The  interval $[-y_\mathrm{max},y_\mathrm{max}]$ can be adjusted to be in agreement with the constraint of light-cone momentum conservation. Since the transverse extension of the volume is fixed, the remaining longitudinal extension $L_z$ of the volume is then fitted to the total multiplicity. We find the following values for the boosted thermal fireball:
\begin{align}
T &=\unit[0.189]{GeV} \\
y_\mathrm{max} &= 4.0 \\
L_z &=\unit[3.4]{fm}
\end{align}
\begin{figure}
\centering
\includegraphics[width=\linewidth]{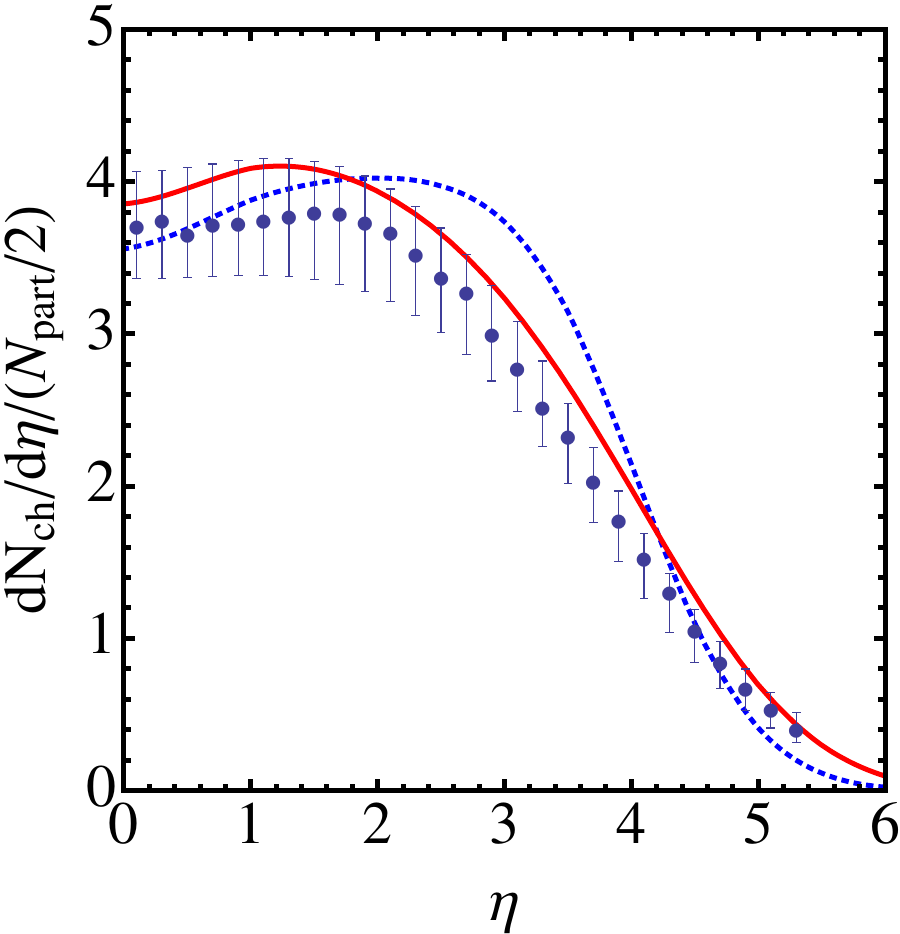}
\caption{Comparison of the boosted fireball distribution (dashed line) and the light-cone distribution (solid line)
satisfying the same constraints (see text) in comparison with data \cite{Alver:2010ck}.}
\label{fig:comp_entropy}
\end{figure}
The boosted thermal distribution may more appropriately describe a series of hadronic fireballs formed later in the collision at a lower temperature. 

In order to compare entropies we substitute our light-cone distribution and the corresponding boosted thermal distribution function into Eq.~\ref{eq:entropy2} for the entropy: 
\begin{align}
n_{x,p_{\bot}} &= \frac{1}{e^{\frac{m_\bot}{\lambda}+x w }-1} \\
n^\mathrm{bt}_{x,p_{\bot}} &= \frac{L_z}{2\pi}
\int_{-y_\mathrm{max}}^{y_\mathrm{max}} \mathrm{d}u \frac{m_{\bot} \cosh(y-u)}{\exp\left(m_{\bot} \cosh(y-u)/T\right)-1}.
\end{align}
The impact parameter dependence is homogeneous in an area of size $L_{\bot}^2$. 

We obtain for the light-cone distribution and the boosted thermal distributions rather similar entropy values with a larger entropy for the light-cone distribution as it should be:
\begin{align}
S^\mathrm{lc} &= 142.4\\
S^\mathrm{bt} &= 141.1
\end{align}

This minimal difference does not favor the light-cone distribution strongly.  Comparing the shapes of the rapidity distributions in Fig.~\ref{fig:comp_entropy}, however, one sees that the boosted thermal distribution does not agree as well with the data as the faster-falling light-cone distribution. One should also consider that the smearing function of the thermal distribution is theoretically not fixed and the above choice is arbitrary; in fact, also a Gaussian averaging has been proposed \cite{Becattini:2007ci}. All fireballs at varying rapidities have the same longitudinal extension $L_z$ in spite of boosting: 
\begin{equation}
L_z=V/L_{\bot}^2
\end{equation}
Owing to the light-cone sum rule there are only two parameters in the light-cone distribution. This smaller number of input parameters together with the maximum entropy argument and the better agreement with the data favor the light-cone plasma distribution.  
 
\section{Conclusions}
In regard of the simplicity of the maximum-entropy ansatz, the light-cone plasma distribution is very successful. Minimal experimental information about the mean transverse momentum and the active area of the colliding hadrons optimally account for rapidity and transverse momentum distributions in nucleon-nucleon and nucleus-nucleus collisions. Therefore, many other features of heavy-ion collisions should be reconsidered. The entropy of fireballs distributed uniformly in rapidity is only slightly smaller than the entropy of the light-cone distribution. Conceptually, however, the light-cone distribution presented in this paper emphasizes the non-equilibrium nature of the collision process. Its parameters, an effective transverse temperature and (longitudinal) softness reflect the asymmetry of transverse and longitudinal momenta of the produced particles. In nuclei, multiple scattering of partons leads to an increase of the mean transverse momentum of produced particles which correlates strongly with the central rapidity density. In the future we plan to include different distributions for quarks and gluons and study fluctuations in more detail. This may lead to slight changes of the  rapidity and transverse distributions. Such an extended model would also allow to calculate specific hadronic spectra.

\begin{acknowledgments}
H.J.P. would like to thank J.P. Vary whose encouragements were very important to push the project ahead. We are grateful to J.P. Blaizot and L.~McLerran for helpful comments.
\end{acknowledgments}


\end{document}